\documentclass{article}
\usepackage{emulateapj,amsfonts,psfig}
\def\ergs{\rm \ erg \, s^{-1}}
\def\deg {^\circ}

\def\cmdue {\rm \ cm^{-2}}

\def\ltsima{$\; \buildrel < \over \sim \;$}
\def\lsim{\lower.5ex\hbox{\ltsima}}
\begin{document}
\psfull
\received{~~} \accepted{~~}
\journalid{}{}
\articleid{}{}

\title{The Brera Multi-scale Wavelet (BMW) ROSAT HRI source catalog. \\
II: application to the HRI and first results.}

\author{Sergio Campana, Davide Lazzati\altaffilmark{1},
Maria Rosa Panzera, Gianpiero Tagliaferri}

\affil{Osservatorio Astronomico di Brera, Via Bianchi 46, I-23807
Merate (Lc), Italy}

\altaffiltext{1}{Dipartimento di Fisica, Universit\`a degli Studi di Milano,
Via Celoria 16, I-20133 Milano, Italy}

\submitted{Accepted for publication in ApJ, main journal}

\begin{abstract}
The wavelet detection algorithm (WDA) described in the accompanying paper by 
Lazzati et al. is made suited for  a fast and efficient analysis of images 
taken with the High Resolution Imager (HRI) instrument on board the ROSAT 
satellite.
An extensive testing is carried out on the detection pipeline:
HRI fields with different exposure times are simulated and analysed in the 
same fashion as the real data. Positions are recovered with few arcsecond 
errors, whereas fluxes are within a factor of two from their input 
values in more than 90\% of the cases in the deepest images. 
At variance with the ``sliding-box'' detection 
algorithms, the WDA provides also a reliable description of the source extension,
allowing for a complete search of e.g. supernova remnant or cluster of galaxies 
in the HRI fields.
A completeness analysis on simulated fields shows that for the deepest 
exposures considered ($\sim 120$ ks) a limiting flux of $\sim 3\times 
10^{-15}\ergs\cmdue$ can be reached over the entire field of view.
We test the algorithm on real HRI fields selected for their crowding 
and/or presence of extended or bright sources (e.g. cluster of galaxies 
and of stars, supernova remnants).
We show that our algorithm compares favorably with other X--ray detection 
algorithms such as XIMAGE and EXSAS.

The analysis with the WDA of the large set of HRI data will allow to survey 
$\sim 400$ square degrees down to a limiting flux of $\sim 10^{-13}\ergs\cmdue$ and 
$\sim 0.3$ square degrees down to $\sim 3\times 10^{-15}\ergs\cmdue$. 
A complete catalog
will result from our analysis: it will consist of the Brera Multi-scale Wavelet
Bright Source Catalog (BMW-BSC) with sources detected with a significance 
$\gtrsim 4.5\,\sigma$ and of the Faint Source Catalog (BMW-FSC) with sources 
at $\gtrsim 3.5\,\sigma$.
A conservative estimate based on the extragalactic $\log(N)-\log(S)$ indicates that 
at least 16000 sources will be revealed in the complete analysis of the whole HRI 
dataset.
\end{abstract}


\section{INTRODUCTION} 

The next generation of X--ray instruments, such as AXAF or XMM, 
will provide deep X--ray images with very high source density  
(up to $1000/{\rm deg{^2}}$). 
To fully exploit the scientific content of these data, new and more refined 
detection techniques have to be considered.
Algorithms based on the wavelet transform provide one of the best 
analysis tools, which has been already used in astronomy over the
last decade (Coupinot et al. 1992; Slezak et al. 1993, 1994; Rosati 1995; 
Rosati et al. 1995; Grebenev et al. 1995; Damiani et al. 1997a; Vikhlinin et al. 
1998).

We have fully implemented a Wavelet Detection Algorithm (WDA) in order to meet 
the confidence requirements needed to deal with a large set 
of data (see the accompanying paper by Lazzati el at. 1999; Paper I hereafter).
Here we focus on the application of this WDA to the X--ray images 
taken with the High Resolution Imager (HRI) on board the ROSAT satellite, 
detailing the HRI-specific features of the algorithm and presenting the 
first results of an on-going automatic analysis on all available data.

Catalogs of X--ray sources with more than thousands of objects
have been produced in the last few years (WGA: White, Giommi \& Angelini 1994; 
ROSATSRC: Zimmermann 1994; RASS: Voges et al. 1996; ASCASIS: Gotthelf \& White 
1997; ROSAT Results Archive: Arida et al. 1998).
These catalogs are mainly based on the Position Sensitive Proportional Counter 
(PSPC) on board the ROSAT satellite and have been heavily used 
in many research projects on different class of X--ray sources
(e.g. Padovani \& Giommi 1996 on blazars; Fiore et al. 1998 on quasars; 
Angelini, Giommi \& White 1996 on X--ray variable sources; Israel 1996 
on X--ray pulsars).

The most appealing feature in using the ROSAT HRI data rather than the PSPC one
is provided by the sharp core of the Point Spread Function (PSF), 
of the order of few arcseconds FWHM on-axis. 
This allows to detect and disentangle sources in very crowded 
fields and to detect extended emission on small angular sizes. 
Moreover, the search for counterparts at different wavelengths will be greatly 
simplified by the reduced error circles.
On the other hand, the ROSAT HRI instrument has a very crude spectral 
resolution, thus a spectral analysis can not be carried out. 
It is less efficient than the PSPC by a factor of 
$\sim 4$ (3--8 for a plausible range of incident spectra) and, finally, 
it has a higher instrumental background.
More details on the performances of the HRI are given in Section 2.
In Section 3 we describe the application of the WDA to HRI data.
Section 4 is devoted to the illustration of the simulations
carried out to test the detection pipeline.
In Section 5 we show the WDA results on ``difficult'' fields (e.g. very 
crowded fields, clusters of galaxies and of stars, supernova remnants).
In Section 6 we summarise our conclusions. 

\section{HIGH RESOLUTION IMAGER}

The HRI on board the ROSAT satellite is a position sensitive detector 
based on microchannel plates, that reveals single X--ray photons 
and determines their positions and arrival times (for more details 
see David et al. 1998). 
The ROSAT HRI is very similar to the Einstein HRI. The nominal pixel size
of $0.5''$ has been reduced to $0.4986''$ after detailed observations on the 
Lockman hole field (Hasinger et al. 1998).
The HRI field of view is given by the intersection of the circular detector
and a square readout, resulting in octagon-like shape with $\sim 19'$ radius.
The HRI PSF on-axis is $\sim 5''$ FWHM, well  
modeled with two Gaussians plus an exponential function. Due to random errors in 
the aspect solution however, images may be occasionally ellipsoidal 
and the PSF parameters may vary up to $\sim 15\%$.
The HRI PSF degrades rapidly for off-axis angles beyond $\sim 5'$ whereas
it becomes very asymmetric beyond $\sim 12'$.
The PSF off-axis is not known with good accuracy and, up to now, only 
a description of the Gaussian widths with the off-axis angle has been 
published (David et al. 1998). Here we adopted a different approach. 
We make use of a ray-tracing simulator in order to extrapolate the  
well known on-axis PSF to any off-axis angle; in this way we get rid of 
aspect solution and photon statistic problems (see Appendix A for more detail).

The on-axis effective area is 83 cm$^2$ at 1 keV, while the vignetting is less 
than 10\% within $10'$ at all energies. The effective area has not 
varied significantly since launch.
Systematic uncertainties amount to $\lesssim 3\%$.

The HRI covers an energy range of 0.1--2.4 keV, divided 
in 16 Pulse Height Analyzer (PHA), which provide very crude spectral information. 
Hardness ratios can give some qualitative information, but the 
gain variations lead to the definition of the PHA boundaries on a case by 
case basis (Prestwich et al. 1998).

The HRI background is made of several components: the internal background 
due to the residual radioactivity of the detector (1--2 cts s$^{-1}$), 
the externally-induced background from charged particles (1--10 
cts s$^{-1}$) and the X--ray background (0.5--2 cts s$^{-1}$).
The background is the highest in the first few (1--3) PHA channels and it
is dominated by the first two components. As shown by sky calibration sources, most of 
the source photons instead fall between PHA channel 3 and 8, approximately
(David et al. 1998).

\section{Wavelet Detection Algorithm for the HRI}

\begin{figure*}[!htbp]
\centerline{\psfig{figure=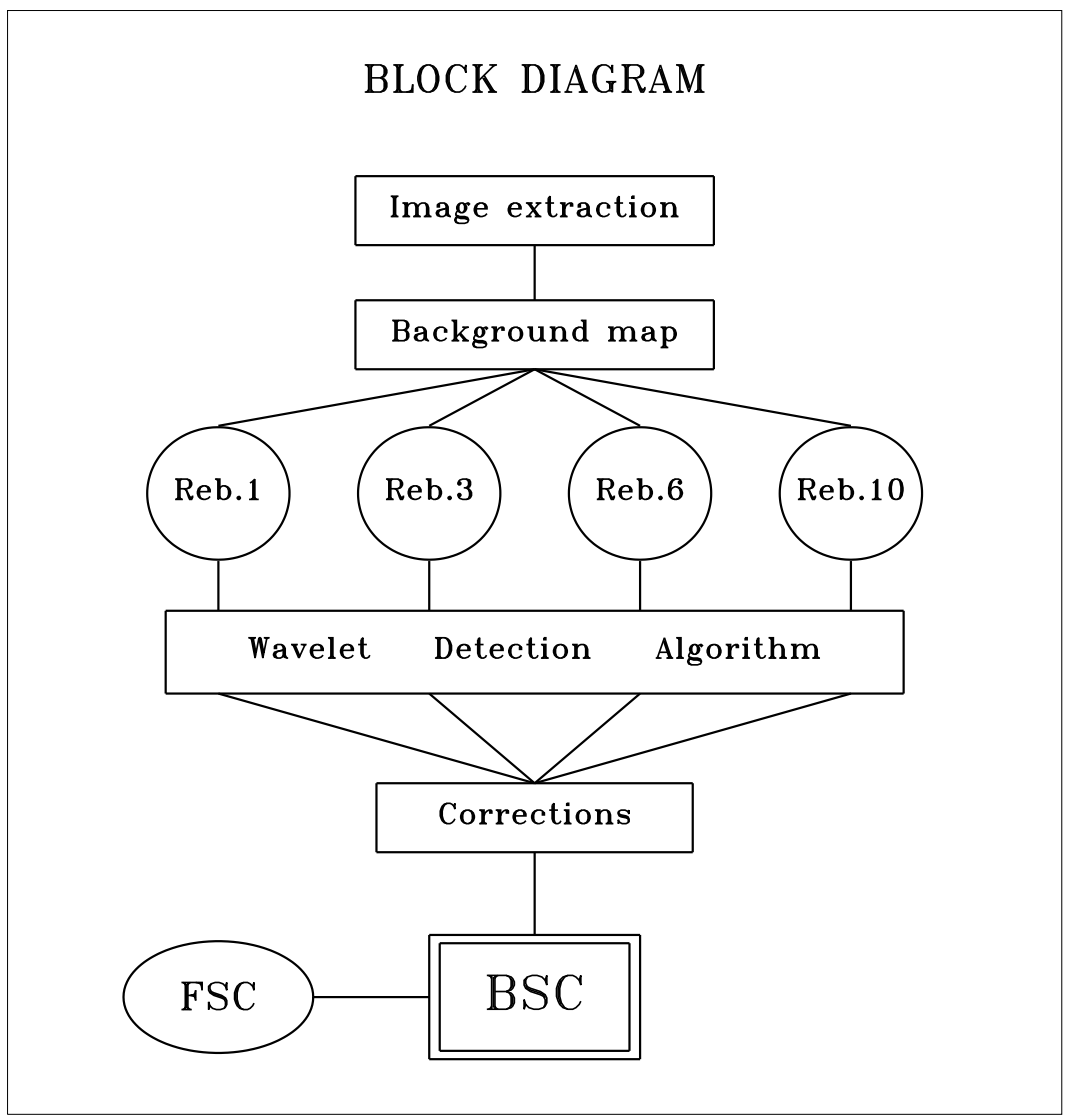,width=14cm}}
{\caption{Block diagram of the BMW pipeline.}
\label{block}}
\end{figure*}

The analysis and source detection of HRI images takes place in several
steps, here we briefly describe the analysis of HRI images 
(see Fig.~\ref{block}). The detection algorithm is presented and described 
in detail in Paper I (see also Fig.~1 therein).

\subsection{Image extraction}

Due to computer limitations it is not efficient  to analyse the entire HRI
image in one single step, preserving the original angular resolution
(it would result in a $\sim 5000\times 5000$ pixels image).
In order to maintain the superb HRI angular resolution, we analyse the 
images in more steps. We extract a $512\times512$ pixels image rebinned by a 
factor of 1 (pixel size $0.5''$), as well as concentric images rebinned by a 
factor of 3 (pixel size $1.5''$), 6 (pixel size $3''$) and 10 (whole image with 
pixel size $5''$), respectively.
These images are extracted from the relevant event files using the XIMAGE 
program (Giommi et al. 1992). Each image is then analysed separately.

\subsection{Background estimate}

For each image a mean background value is estimated using a $\sigma-$clipping 
algorithm 
(see Paper I): the mean and standard deviation of an image are calculated and 
pixels 
at more than $3\,\sigma$ above the mean value are discarded.
The procedure works iteratively, rebinning the image and discarding outstanding 
pixels. We carried out various simulations on HRI fields and found that 
even in crowded fields the background value is recovered within
$\sim 10\%$. We checked these values using the background estimator 
within XIMAGE obtaining the same results and accuracy.

\subsection{Exposure map}

X--ray mirrors, like optical mirrors, are vignetted.
This generates non-flat fields where the detection of X--ray sources 
is made difficult by the second-order derivative of 
the background component (the adopted WT is insensitive to constant or 
first-order background components, see Paper I). In addition to this 
component, there may be obscuring structures (like the ROSAT PSPC rib) 
or hot regions in the detector (like in the EXOSAT Channel Multiplier Array)
or real sky background variations (e.g. in the presence of extended sources).
For these reasons it is usually better to perform the source search 
on top of a background map, rather than on a flat background.
In the particular case of the ROSAT HRI, this allows us to search also
for sources which lie in the edge region, where the detector efficiency 
rapidly drops to zero.
In this case the linearity of the wavelet transform helps us:
since an image can be thought as the sum of a background component
plus the sources, the transform of the source component can be obtained 
subtracting the transform of the background map from that of the whole image.
Background maps are provided together with the relevant data 
(as in the case of ROSAT PSPC) or can be generated through dedicated 
software (as the ESAS software for both ROSAT PSPC and HRI instruments; 
Snowden et al. 1994). 

A different approach is to properly smooth
the source image, filtering out the brightest sources, and using this 
image as the background map (Vikhlinin et al. 1995b; Damiani et al. 1997a).
This approach is also useful in the presence of extended emission which cannot 
be modeled analytically (e.g. supernova remnants) even if the smearing
of extended and faint sources tends to reduce source significance and 
hence the completeness of the catalog.

At variance with the ROSAT PSPC, the HRI background is dominated by the 
unvignetted particle background. In order to minimise the impact of 
this background and more generally to increase the signal-to-noise (S/N)
ratio of X--ray sources, we restrict our analysis to PHA 
2--9 (see also David et al. 1998; Hasinger et al. 1998). 
This range reduces the detector background by about 40\%
with a minimum loss of cosmic X--ray photons ($\lesssim 10\%$; David et al. 1998). 
To build the exposure map we adopt the ESAS software (Snowden 1994)
that makes use of the bright Earth and dark Earth data sets
to produce a vignetted sky background map and a background detector map, 
respectively. The mean image background, as estimated with the $\sigma-$clipping algorithm 
(see above), is used to normalise the sky background map provided by the ESAS software.
The total background map is then obtained by summing up the detector and sky maps.
The ESAS maps are produced at rebin 10 (pixel scale $5''$), 
thus we interpolate them to obtain the maps at rebin 1, 3 and 6. 
These three maps are then smoothed with a Gaussian filter  
with a size twice the mean PSF, in order to get rid of interpolation 
inhomogeneities.

\subsection{Image analysis}
 
The four images and relative exposure maps are searched for significant enhancements 
using the WDA. 
On the first three images the detection is performed within an annulus of 
255 pixels excluding a border at the detector edge of the local PSF width
($\sim 5''$, $\sim 7''$ and $\sim 12''$ at rebin 1, 3 and 6, respectively). 
This strategy has been adopted in order to 
preserve the original circular symmetry of the image and to avoid the occurrence 
of azimuthally dependent detection thresholds.
In fact, detecting sources on the whole $512\times 512$ pixels image would result 
in a bias at the corners between 255 pixels (i.e. the image radius) 
and $255\,\sqrt{2}$ (i.e. half the square diagonal). 
Sources in the region left over are recovered at the successive rebin. At rebin 
10 the whole image is handled.

The detection threshold choice has deep impact on the characteristics of 
the catalog to be produced. 
The use for statistical purposes (such as $\log(N)-\log(S)$) of the catalog requires
a high threshold (i.e. low contamination), whereas a low threshold is needed for 
the detection of a large number of sources, even if plagued by higher contamination.
For these aims, we consider two different detection thresholds: a contamination of 
0.1 spurious source per field is allowed for the Brera Multi-scale Wavelet Bright 
Source Catalog (BMW-BSC); a contamination of 1 spurious source per field is used for the 
Faint Source Catalog (BMW-FSC). The equivalent thresholds applied to 
BMW-BSC and BMW-FSC correspond to a source significance of $\sim 4.5\,\sigma$ and 
$\sim 3.5\,\sigma$, respectively.
These detection thresholds are given for each field as a whole and must 
therefore be shared among the four rebinned images, holding the relative 
contamination constant over the full HRI field. 
The detection threshold for each rebinned image cannot be just one forth 
of the whole threshold (e.g. 0.25 in the case of 1 spurious source per field), 
due to the different area analysed at each rebin.
A proper weighting factor is given by the ratio between the analysed area 
and the mean PSF width at each rebin.

\subsection{Corrections}

The HRI data are pre-processed with the ROSAT Standard Analysis System 
(Voges 1992) which provides images corrected for detector non-linearities 
and attitude control.
We also apply the vignetting and gain correction, deriving them from the 
final exposure map. A deadtime correction in the case of bright sources 
is also applied (David et al. 1998).

To these we have to add other small corrections related to the detection 
algorithm we use.
To fit the sources in the wavelet space we approximate the PSF with a single
Gaussian. However, the PSF has an extended tail and becomes increasingly asymmetric 
with the off-axis angle, thus we systematically loose counts.
Since the HRI PSF is not well known at large off-axis angles, we 
performed ray-tracing simulations all over the field of view (see Appendix A). 
We then fit these ray-tracing simulated images with the WDA and derived the PSF 
correction needed to obtain the real flux (the PSF correction is a common 
characteristic of all detection algorithms). 
This PSF correction depends on the off-axis angles 
and varies from $\sim 18\%$ for sources at off-axis angles $\theta\lesssim 2'$
(i.e for all sources in the image at rebin 1) to $\sim 13\%$ for 
$6'\lesssim \theta\lesssim 13'$ (i.e. at rebin 6; see Table \ref{cts_off}).
At rebin 10 the PSF degradation makes this correction a steep function of 
the off-axis angle (see Table \ref{cts_off}). The reduction of the PSF correction 
with the off-axis angle is likely due to the vanishing importance of the second 
Gaussian in the PSF.

\begin{figure*}[!htbp]
\centerline{\psfig{figure=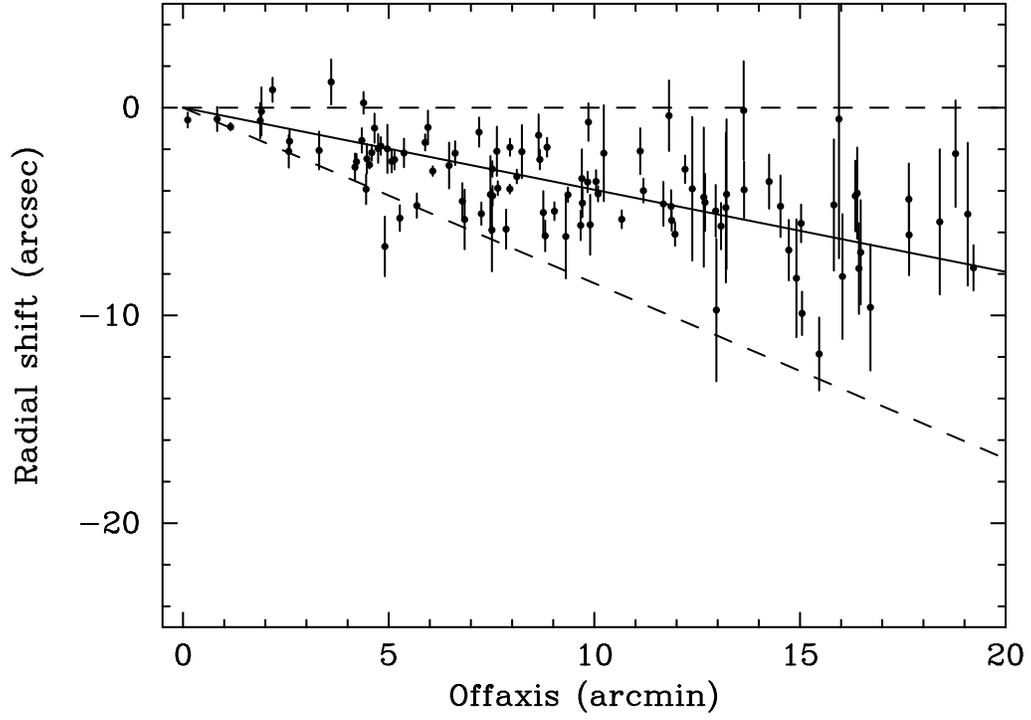,width=10cm}}
{\caption{Radial shift between the X--ray source positions and their GSC 
optical counterparts versus the off-axis angle. The continuous line 
represents the best fit to the data. 
Dashed lines mark a conservative $3\,\sigma$ limit to the fit
(see text).}
\label{off}}
\end{figure*}

The HRI PSF becomes increasingly asymmetric with the off-axis angle, developing
a bright spot (the center of which coincides with the source position) on top 
of an extended emission (as shown also by our ray-tracing simulations). The center  
of this extended emission is shifted by a few arcsec from the spot in the  
direction opposed to the field center. 
Fitting the counts distribution of an X--ray source with a Gaussian, the WDA 
finds its position in between the bright spot 
and the center of the extended emission (even if much nearer to the spot).
In order to correct for this (small) effect which is mainly due to the larger 
support of the wavelet functions, we selected 4 HRI exposures with a 
large number of sources (Trapezium ROR 200500; P1905 ROR 
200006; IC348 ROR 201674 and NGC2547 ROR 202298). 
We first performed a boresight correction on the central sources, then we 
selected X--ray sources associated with Guide Star Catalogue objects 
within $10''$.
We measured the distance between the X--ray sources and the optical counterparts 
as a function of the off-axis angle.  
Even if a (small) number of spurious identifications can take place, we note
that a systematic effect is clearly evident in Fig.~\ref{off}.
In particular the radial shifts are linearly correlated with the off-axis angle 
and the best fit line provides a correction of about $7''$ for a source at $18'$ 
off-axis (this value includes also the $\sim 3''$ correction for the smaller 
pixel size, cf. Section 2).
The scatter in the source off-sets is rather large so we decided to 
conservatively compute the $3\,\sigma$ uncertainty on the best fit line by 
individuating the region that includes the 99.7\% of sources (cf. 
Fig.~\ref{off}). 
This is achieved by considering the two dashed lines in Fig.~\ref{off}. The derived
error is of about $7''$ for a source at $18'$ off-axis at a $3\,\sigma$ level.
This shift error has been summed in quadrature to the position error derived 
from the fit and it provides the total error for the sources in our catalog.

We point out that this result should not be regarded as a different 
HRI plate scale as reported in David et al. 1998 and Hasinger et al. 1998.
The radial shift found in this analysis is mainly caused by the larger support
of the wavelet functions that probe the source PSF on a larger scale than the
sliding box techniques.

\subsection{Creation of the catalog}

For each observation we derive a catalog of sources with
position, count rate, extension along with the relative errors, as well as 
ancillary information about the observation itself and source fitting.
The count rate has been computed by fitting the image transform 
at all scales simultaneously in the wavelet space (see Paper I). 
We provide also a second count rate estimate following the standard approach 
of counting the source photons, however this method fails e.g. in the case 
of crowded fields or extended emission.
The counts to flux conversion factor is determined based on a Crab spectrum.
In the case of high latitude fields ($|b| > 30\deg$) the galactic column 
density is assumed, whereas for lower latitude fields we consider either a 
null column density and the galactic value, therefore providing a range of fluxes.
Together with the information relative to the X--ray data, we cross
correlate the detected sources with databases at different wavelengths 
to give a first identification.

\begin{figure*}[!htbp]
\centerline{\psfig{figure=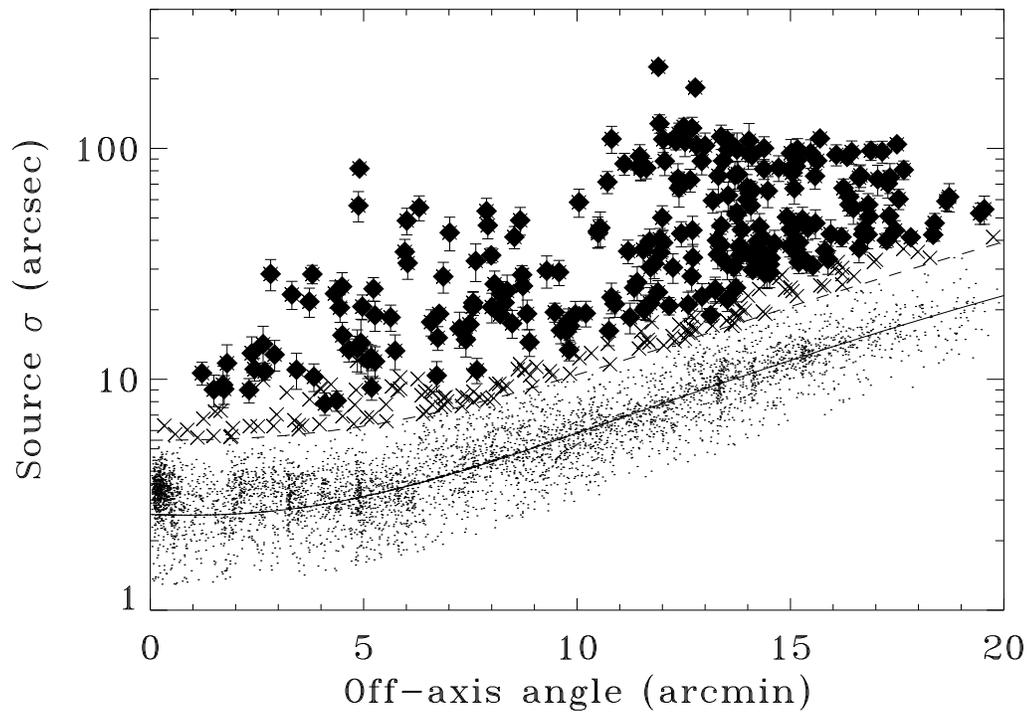,width=15cm}}
{\caption{Source extension ($\sigma$) versus off-axis angle for 6013 sources
detected in HRI fields on stars (BMW-BSC). 
The dashed line marks the $3\,\sigma$ extension limit for point sources,
whereas the continuous line the computed PSF. To be more conservative we 
consider as extended the sources lying more than $2\,\sigma$ above this line 
(filled squares). Crosses marks sources which lie at a significance lower than 
$2\,\sigma$ from this line. 254 sources have been classified as extended.}
\label{extend}}
\end{figure*}

One of the most interesting feature of the wavelet analysis is 
the possibility of characterising the source extension, however this
cannot be assessed simply by comparing the source width 
($\sigma$, as derived by the fitting procedure) with the HRI PSF (e.g. as derived 
by the ray-tracing simulator) at a given off-axis angle, due to the 
energy dependence of the PSF width as well as to errors 
in the aspect reconstruction (near on-axis). 
Thus, to assess the source extension, we considered a version of the 
catalog consisting of sources detected in the observations that have a star(s)
as a target (ROR number beginning with 2). 
We considered 756 HRI fields and we detected 6013 sources
in the BMW-BSC catalog (Fig.~\ref{extend}).
The distribution of source extensions has been divided into bins of $1'$ each, 
as a function of the source off-axis angles. 
In each $1'$ bin, we applied a $\sigma-$clipping algorithm on the source extension: 
the mean and standard deviation in each bin are calculated and sources with widths
at more than $3\,\sigma$ above the mean value are discarded.
This method iteratively discards truly extended sources and provides
the mean value of the source extension ($\sigma$) for each bin along with its error.
We then determine the $3\,\sigma$ dispersion on the mean extension for each 
bin. The mean value plus the $3\,\sigma$ dispersion provide the line demarking 
source extension (cf. the dashed line in Fig.~\ref{extend}; see also Rosati et al. 1995). 
We conservatively classify a source as extended if its error on the extension 
parameter is such that it lays more than $2\,\sigma$ from this limit 
(see Fig.~\ref{extend}). Combining this threshold with the $3\,\sigma$ on the 
intrinsic dispersion we obtain a $\sim 4.5\,\sigma$ confidence level for the 
extension classification.
254 sources have been classified as extended, which makes up $\sim 4\%$ 
of the total. Note that no source has been classified as extended on-axis, 
as should be expected being stars the targets.

A word of caution has to be spent for the flux estimate of extended sources.
The source flux is computed by fitting a Gaussian to the surface brightness profile 
and, in many cases, this provides a poor approximation. Therefore, fluxes of 
extended sources are usually underestimated. A solution in the case of extended 
sources with well-defined surface brightness profiles (i.e. clusters of galaxies)
has been presented by Vikhlinin et al. 1998.

\medskip
\section{SIMULATIONS}

The WDA presented in paper I has provided very good results on ideal fields,
with flat background, Gaussian sources and no crowding.  
To test the whole pipeline also on realistic images,
we simulated sets of HRI observations using real instrumental 
background maps generated with the Snowden's software and superposing 
X--ray sources following the soft X--ray $\log(N)-\log(S)$ by Hasinger 
et al. 1993 (see also Vikhlinin et al. 1995b).
We took the exposure map of one of the longest observations in the ROSAT public 
archive (NGC 6633 - ROR 202056a01 - $\sim 120$ ks), and 
superpose X--ray sources down to a flux of $\sim 5\times 10^{-16}\ergs\cmdue$ 
(a conversion factor of $1\,{\rm cts\,s^{-1}} = 1.71\times 10^{-11}\ergs\cmdue$ in the
0.5--2.0 keV energy band has been adopted; Hasinger et al. 1998). 
Each simulated image contains 500 sources, the faintest of which have 
$\sim 3$ counts and enhance the sky background being well below the detection 
threshold.
The sources were distributed homogeneously all over the field of view 
and the appropriate PSF obtained with the ray-tracing simulator was 
used to spread their photons.
A total of 100 fields were simulated and analysed in the same automatic 
fashion as the real data. 
Every detected output source has been identified with a simulated 
input source within the $3\,\sigma$ error box. 

\begin{figure*}[!htbp]
\centerline{\psfig{figure=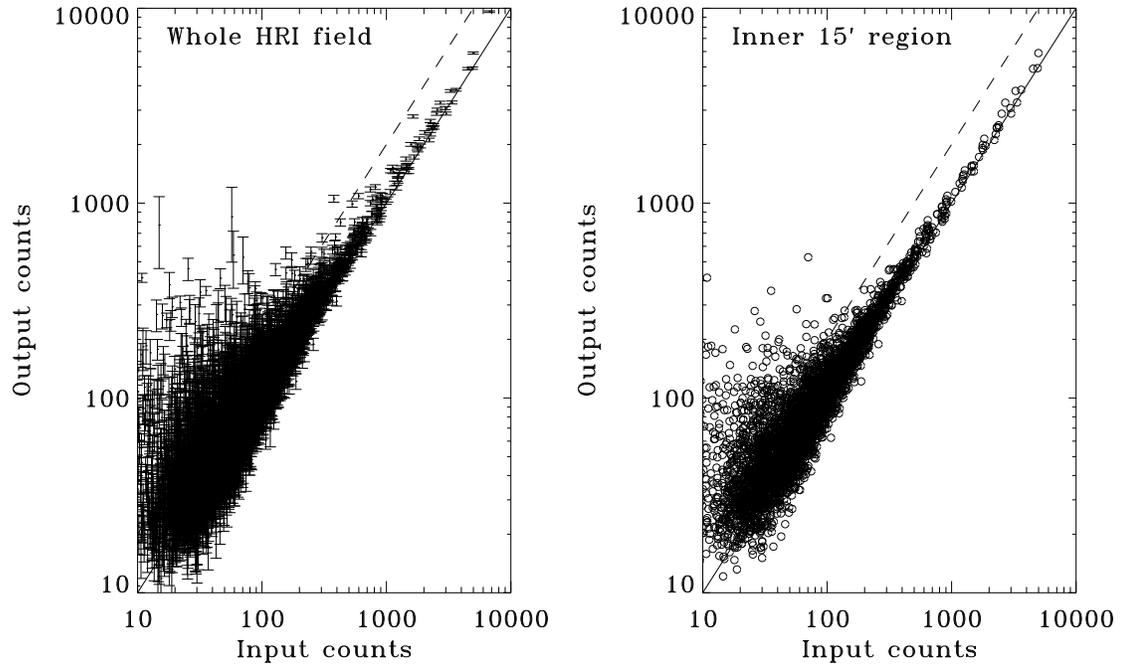,width=15cm}}
{\caption{Input versus output counts for detected sources in 100 simulated 
HRI images (120 ks). The whole field of view has been considered (left panel), 
whereas the results for the innermost $15'$ are shown in the right panel.
More than 90\% of sources have a detected count rate within a factor of two 
the input values over the entire field of view (dashed line).}
\label{cts}}
\end{figure*}

These simulations allow to probe the WDA algorithm behaviour on ``real'' HRI fields, 
revealing the presence and the influence of biases and selection 
effects (e.g. Hasinger et al. 1993; Vikhlinin et al. 1995a). 
The great majority (more than 90\%) of source fluxes are recovered within a 
factor of 2 of their input values.
The tail of the counts distribution starts enlarging at $\sim 120$ counts
over the entire field of view (i.e. $\sim 2\times 10^{-14}\ergs\cmdue$,
see Fig.~\ref{cts}). This effect is produced by the combination of 
source confusion (e.g. Hasinger et al. 1998) and the bias
in the source intensity determination extensively discussed by Vikhlinin 
et al. 1995a. The latter occurs due to the preferential selection of sources 
coincident with positive background fluctuations, near the detection threshold 
and it is more severe for surveys with low S/N ratios.
Comparing the results on the simulated HRI fields with the ones with
equally spaced sources discussed in Paper I (cf. Figure 5), we conclude that 
source confusion is more important. We remark that source confusion affects only 
a small fraction of the sources ($\lsim 10\%$). If, in fact, we quantify the 
confusion following Hasinger et al. 1998 (cf. Equation 5 therein), we are far 
below the strong confusion regime.

\begin{figure*}[!htbp]
\centerline{\psfig{figure=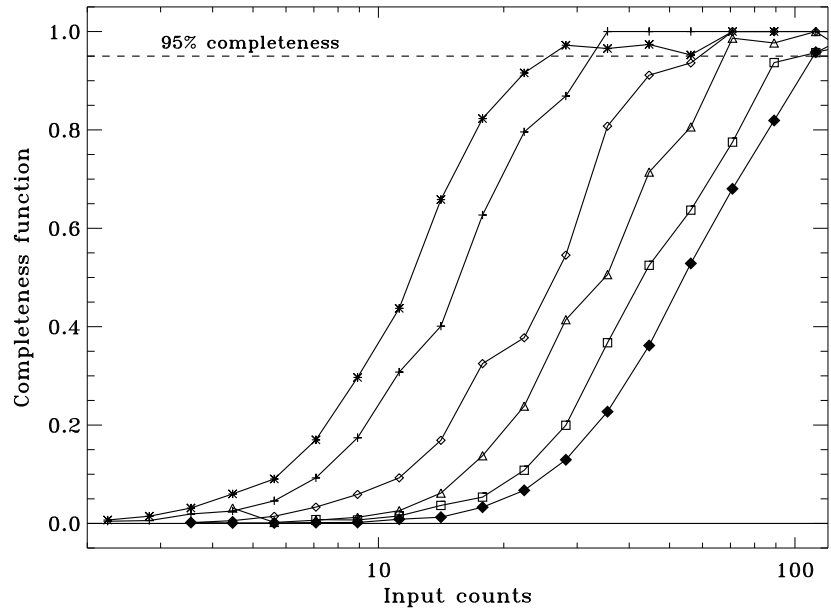,width=12cm,angle=90}}
{\caption{Completeness functions for different exposure times. From left 
to right, the functions refer to simulations of 1, 7, 15, 30, 60 and 120 ks exposure 
time, respectively. The horizontal line shows the 95\% completeness.}
\label{compl_fit}}
\end{figure*}

\begin{figure*}[!htbp]
\centerline{\psfig{figure=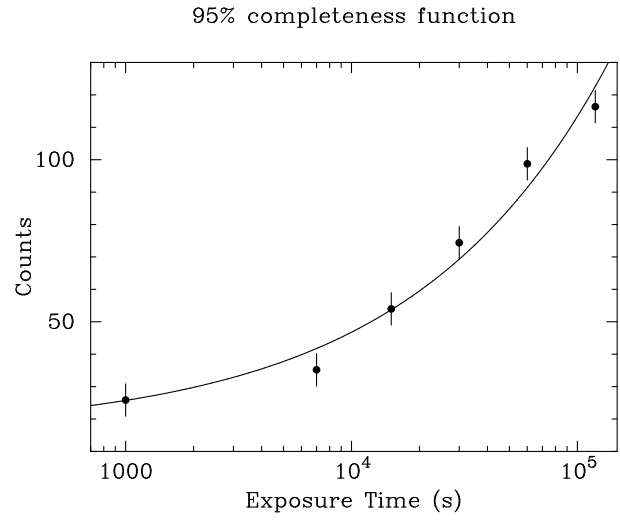,width=10cm,angle=-90}}
{\caption{95\% completeness threshold in total counts as a function of
the exposure time. These values can be modeled by a square root 
function plus a 
constant, showing that the algorithm is background limited 
up to $\gtrsim 100$ ks exposures.}
\label{compl}}
\end{figure*}

In order to explore the influence of biases and selection effects 
we simulated also 50 fields at five different
exposure times (1, 7, 15, 30 and 60 ks) each. Rather than adopting the Hasinger's 
$\log(N)-\log(S)$ distribution (which implies about 3 sources per field at 7 ks), 
we considered in these cases a simpler one, i.e. a power law $\log(N)-\log(S)$ 
with an index of --2.5 and with arbitrary normalisation set to have about 
200 sources per frame. This approach has been adopted in order to reduce the 
number of simulations and therefore computing time (which is mainly spent in the 
calculation of the image and background transforms)
and it results in a more stringent test on the WDA due to the
heavier crowding. The relative completeness functions are
plotted in Fig.~\ref{compl}. As one can see the 95\% limiting flux 
moves from $\sim 4\times 10^{-13}\ergs\cmdue$ to $\sim 3\times 10^{-14}\ergs\cmdue$
(see Fig.~\ref{compl_fit}). These numbers refer to a completeness achieved over
the full field of view. Lower values can be obtained reducing the area of interest.
The total number of counts needed to achieve the 95\% completeness level as a 
function of the exposure time can be well approximated by a constant plus 
a square root function (see Fig. \ref{compl}).

\subsection{General properties of the survey}

\begin{figure*}[!htbp]
\centerline{\psfig{figure=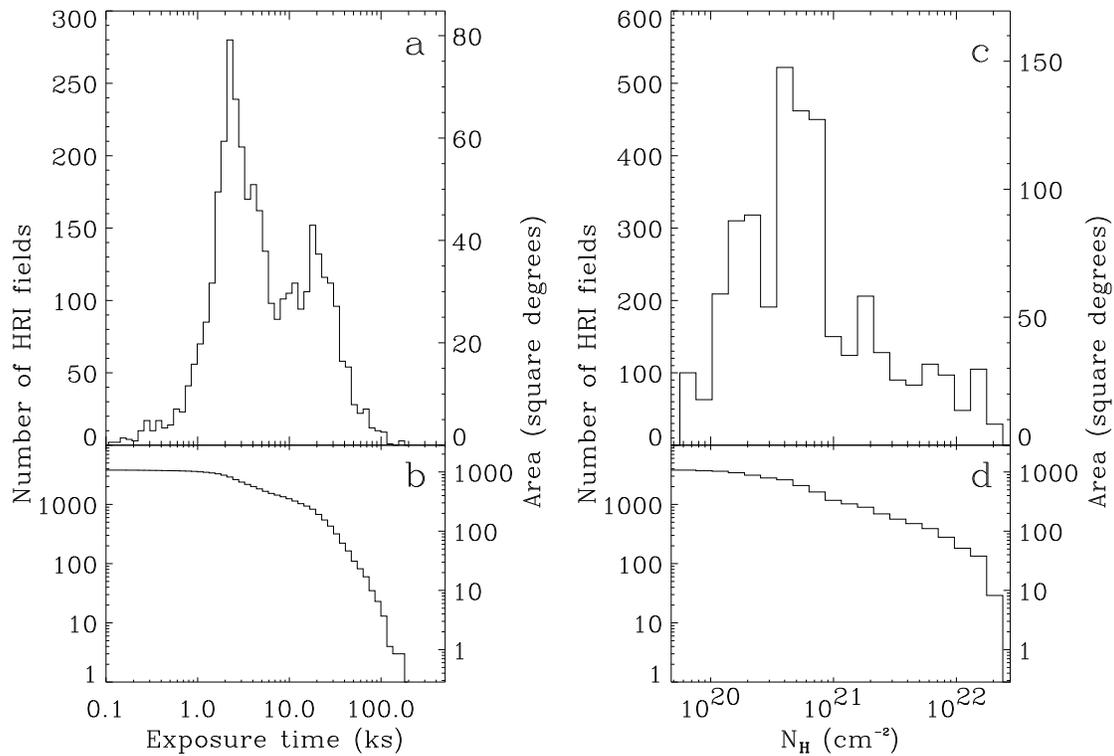,width=15cm}}
{\caption{In the panel ($a$) is shown the distribution of the 
HRI exposure times
and in ($b$) the cumulative distribution. In panel ($c$)
is shown the distribution of the galactic column density and in ($d$) the 
cumulative distribution.}
\label{sky}}
\end{figure*}

The sensitivity of the HRI instrument is not uniform over the field of view 
but decreases rapidly with the off-axis angle as a consequence of the PSF 
broadening and mirror vignetting.
For this reason the surveyed region at a given limiting flux should not
necessarily coincide with the HRI detector area but is in general a smaller 
circular area.
We compute the sky coverage of a single observation by calculating the detection 
thresholds over the entire field of view. With the help of the ray-tracing program we 
simulated sources from $0'$ to $18'$ off-axis with a $1'$ step. 
We then properly rebin these simulated images and calculate the minimum number of 
counts needed to reveal the source at the selected off-axis angle as a function of 
the image background. 
A sky survey of $\sim 400$ square degrees is expected down to a limiting 
flux of $\sim 10^{-13}\ergs\cmdue$, and of 0.3 square degrees down to 
$\sim 3\times 10^{-15}\ergs\cmdue$. 
A preliminary and conservative estimate based on the extragalactic soft X--ray 
$\log(N)-\log(S)$ indicates that about 16000 sources will be revealed in the 
complete analysis of the whole set of HRI data (BMW-BSC). 
In Fig.~\ref{sky} we show the differential and integral distributions of exposure 
times and galactic absorptions.

\begin{figure*}[!htbp]
\centerline{\psfig{figure=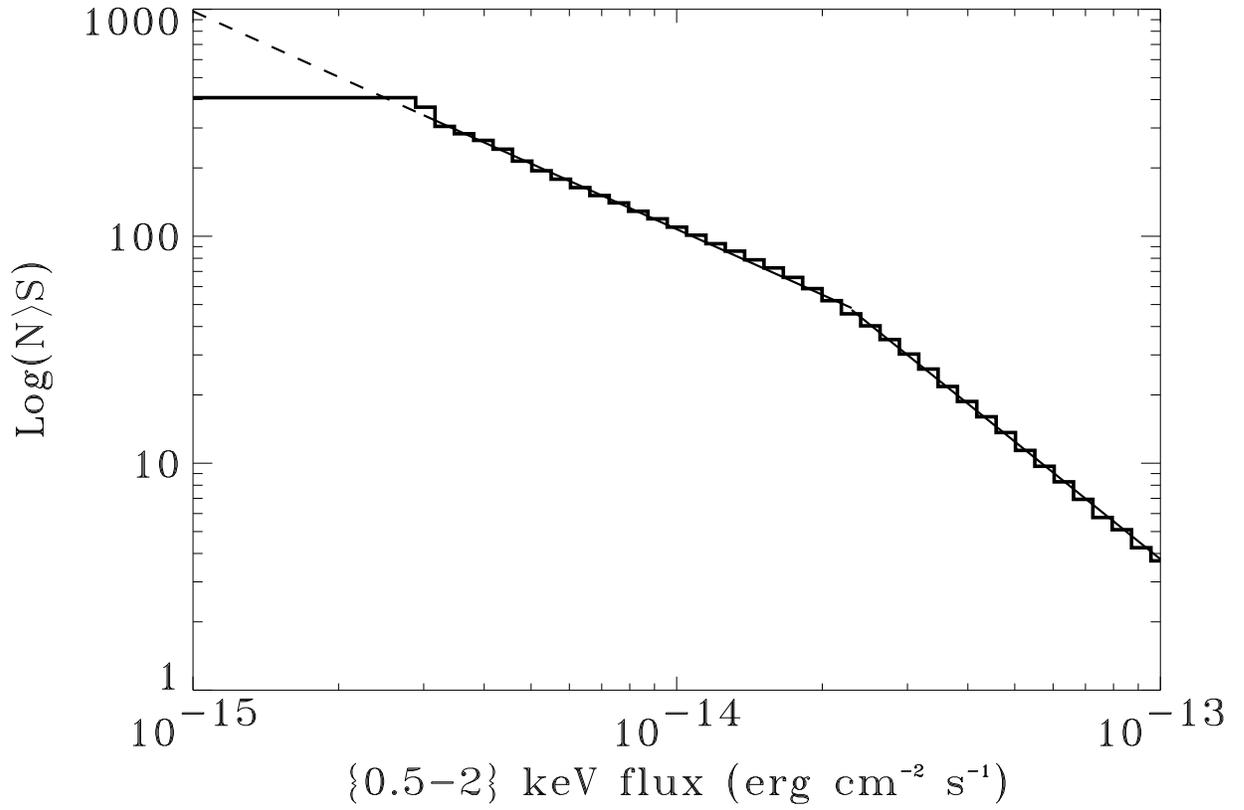,width=17cm}}
{\caption{Measured $\log(N)-\log(S)$ from the analysis of 100 simulated HRI 
fields (120 ks). The continuous-dashed line marks the input distribution of 
sources.
The derived distribution reproduces the input $\log(N)-\log(S)$ down to a 
flux of $\sim 3\times 10^{-15}$ 
erg cm$^{-2}$ s$^{-1}$. A flux to count rate conversion factor of $1\,{\rm 
cts\,s^{-1}}= 1.71\times 10^{-11}$ erg cm$^{-2}$ s$^{-1}$ in the 0.5--2.0 keV
energy band has been adopted.}
\label{lognlogs}}
\end{figure*}

We also computed the $\log(N)-\log(S)$ distribution based on the dataset of 120 ks 
simulations (see above). The recovered distribution is complete down to a flux
of $\sim 10^{-14}\ergs\cmdue$ over the entire ($18'$) field of view. The knowledge 
of the sky coverage allows us to correct for the loss of sources, enabling us
to recover input $\log(N)-\log(S)$ down to a flux of $\sim 3\times 
10^{-15}\ergs\cmdue$ (Fig.~\ref{lognlogs}).

\section{FIRST RESULTS ON SELECTED HRI FIELDS}

To test our HRI pipeline, we examined a sample of HRI fields (see Table \ref{fields})
selected for their ``difficulty'' in terms of source confusion and extended 
emission, for which sliding box techniques face serious problems.
We compare the results obtained with our WDA
with the detections made with XIMAGE/XANADU (Giommi et al. 1992) and 
EXSAS/MIDAS (Zimmermann et al. 1993).
We have to point out that the detection algorithm in XIMAGE is not optimised 
for the HRI so that, especially at large off-axis angles where the PSF
degrades, strong sources are usually detected as multiple;
on the other hand EXSAS/MIDAS has been specifically developed to deal with  
the ROSAT data. 

We ran our WDA on these fields automatically without particular settings.
For XIMAGE we set the probability threshold to $7\times 10^{-6}$ and for EXSAS
the maximum likelihood (ML) to 8 but kept sources with a final ML$\,\ge 12$, 
which correspond to a statistical significance of $\sim 4.5\,\sigma$.

In Table \ref{fields} we report the number of sources detected with the 
different algorithms as well as the number of sources for each rebin image
(in the case of WDA we report the number of sources for each image, rather than in 
a circle as described above, in order to allow 
the comparison).
As one can see the WDA is more efficient in detecting faint sources. 
Even for fields where the number of sources detected with the other two
algorithms is higher than for the WDA, a closer inspection reveals that 
this is typically due to the presence of a strong or extended source, that has 
been splitted into multiple sources.

\begin{figure*}[!htbp]
\centerline{\psfig{figure=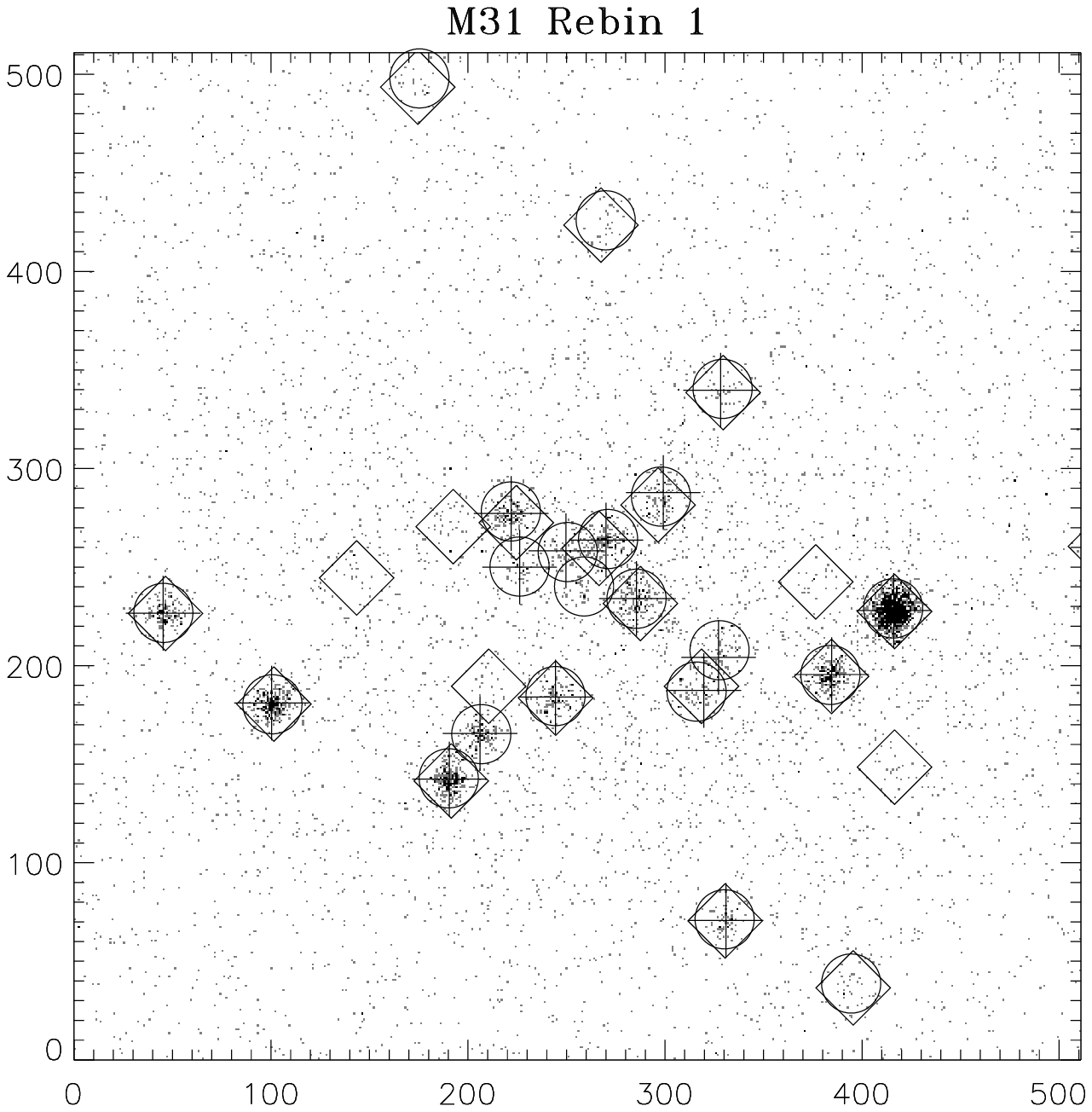,width=10cm}
\psfig{figure=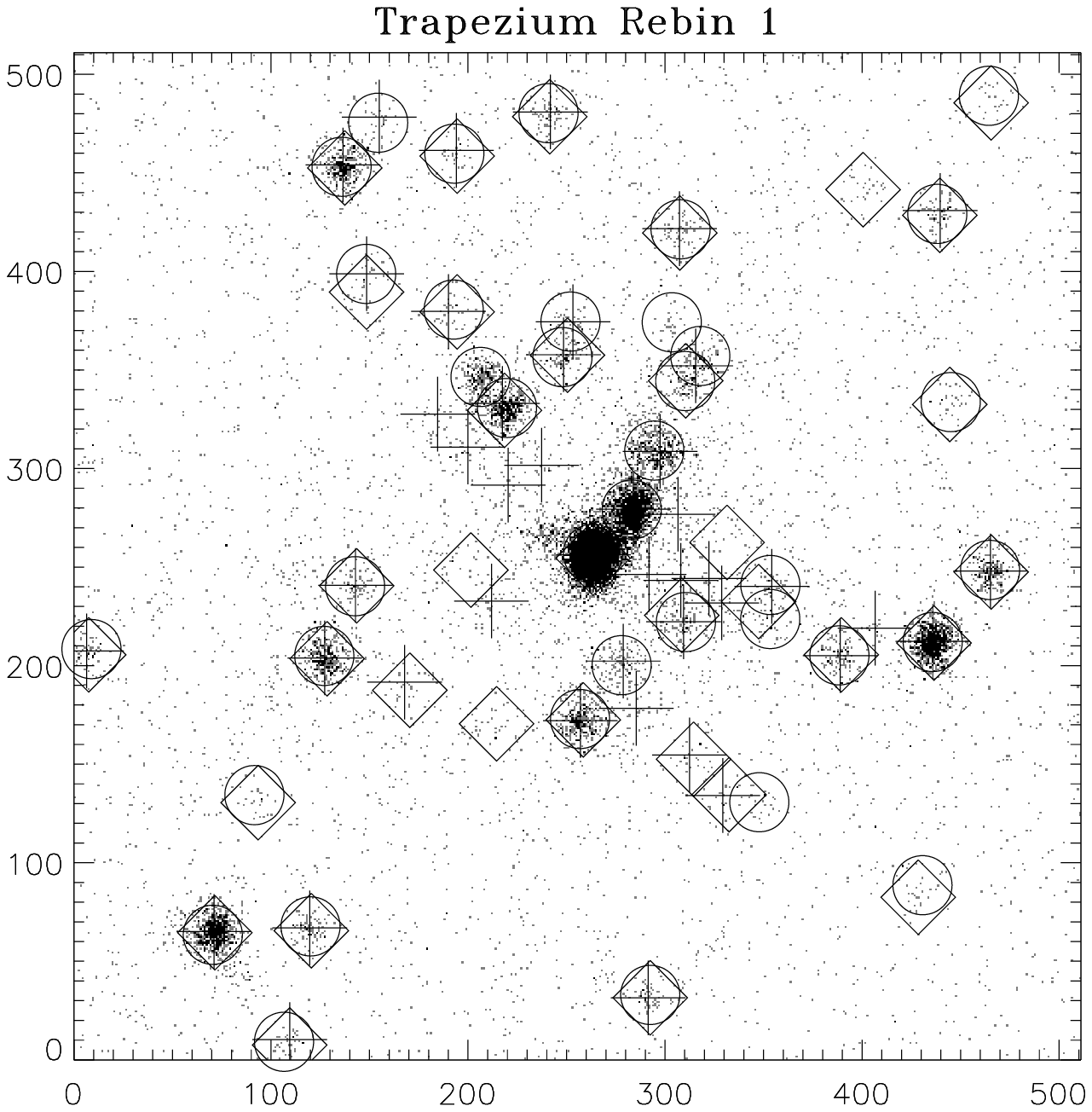,width=10cm}}
{\caption{Source detection in the inner part of the M31 field (left) and of 
the Trapezium star forming region (right). 
Sources detected by our WDA are marked by a circle, those by EXSAS with a 
diamond and those by XIMAGE with a cross.}
\label{m31}}
\end{figure*}

To better compare the algorithms we plot in Fig.~\ref{m31} the inner part 
of the M31 and Trapezium images: of the 59 sources detected (21 in M31 and 38 
in Trapezium) at rebin 1 by our WDA, 35 (13+22) are in common with the other two 
algorithms, 11 (3+8) are found by WDA and EXSAS and 10 (4+6) by WDA and XIMAGE. 
Two sources (0+2) are found by XIMAGE and EXSAS and not by the WDA.
3 (1+2) sources are found by WDA only, 9 (5+4) by EXSAS only and 12 
(0+12) by XIMAGE only. 
If we retain sources detected by at least two algorithms as ``real'' and
sources with only one detection as ``non-real'', we have that the WDA is 
characterised by the smallest number of ``missed'' and ``spurious'' sources.

\section{CONCLUSIONS}

The general theory underlying the WT-based algorithm we developed has been 
described in Paper I, together with the extensive testing we carried out.
Here we focused on the application of this WDA to ROSAT HRI images, describing 
the HRI-dependent features, the major problems encountered (e.g. the sharp drop 
in the background at the detector edges and the PSF broadening with the off-axis 
angle) as well as the extensive testing on simulated images. 
The use of the Snowden's background maps and the modeling of the HRI 
PSF with a ray-tracing simulator allowed us to overcome them and to perform 
the source search over the entire HRI field of view.
In particular, we were able to optimise our WDA such that more than 90\% of 
sources have output fluxes within a factor of two of their input values in 
the deepest images we simulated (120 ks). The use of a wavelet-based algorithm allows 
to flag extended sources in a complete catalog of X--ray sources. 

The completeness functions for different exposure times have been computed
with simulated images. For an exposure time of 120 ks we reach a completeness 
level of 95\% at a limiting flux of $\sim 10^{-14}\ergs\cmdue$ over the full 
field of view. Correcting for the sky coverage the $\log(N) - \log(S)$ distribution 
can be extended down to $\sim 3\times 10^{-15}\ergs\cmdue$ over the entire field 
of view.
The analysis with the WDA of the large set of HRI data will allow a sky survey 
of $\sim 400$ square degrees down to a limiting flux of $\sim 10^{-13}\ergs\cmdue$,
and of $\sim 0.3$ square degrees down to $\sim 3\times 10^{-15}\ergs\cmdue$. 
A conservative estimate based on the extragalactic $\log(N)-\log(S)$ indicates 
that at least 16000 sources will be revealed in the complete analysis of the 
whole set of HRI data (BMW-BSC).

The WDA we developed is also tested on difficult fields and it compares favorably 
with other detection algorithms such as XIMAGE and EXSAS, both for what concerns 
the sensitivity to blended and/or weak sources and the reliability of the detected 
sources.

A complete and public catalog will be the outcome of our analysis: it will consist 
of a BMW-BSC with sources detected with a significance $\gtrsim 4.5\,\sigma$ 
and a fainter BMW-FSC with sources at $\gtrsim 3.5\,\sigma$.
All the detected HRI sources will be characterised in flux, size and position and 
will be cross-correlated with other catalogs at different wavelengths (e.g. Guide Star 
Catalog, NRAO/VLA Sky Survey etc.), providing a first identification.
The BSC can and will be used for systematic studies on different class of sources 
as well as for statistical studies on source number counts.
These images and related information would be available through a multi-wavelength 
Interactive Archive via WWW developed in collaboration with BeppoSAX-Science Data 
Center, Palermo and Rome Observatories. The layout of a typical image is displayed 
in Fig.~\ref{tra_tot}.

The WDA used for the analysis of HRI sources can be adapted for future X--ray 
missions, such as JET-X, XMM and AXAF.
The application of wavelet-based detection algorithms to these new generation of 
X--ray missions will provide an accurate, fast and user friendly source detection 
software.

\acknowledgments{
We thank P. Conconi for his help with the ray-tracing simulator.
We acknowledge S. Snowden for providing a beta version of the ESAS software 
in advance of release.
We are grateful to S. De Grandi, P. Giommi and  P. Rosati for useful comments.
This research has made use of data obtained through the High Energy
Astrophysics Science Archive Research Center (HEASARC) provided by
NASA's Goddard Space Flight Center and through the ROSAT data archive maintained 
at the Max Planck Institute f\"ur Extraterrestrische Physik.
This work was supported through CNAA and ASI grants.}

\appendix{}

In order to study the PSF of the ROSAT HRI instrument one should rely on 
calibration images. However due to problems with the image reconstruction, 
only the on-axis PSF has been worked out to date (David et al. 1998).
Even if the HRI PSF is effectively used in the detection algorithm 
only to correct the count rate (i.e. the PSF correction) and for the sky coverage, 
its knowledge is extremely useful also to cross check other results 
(e.g. the radial dependence of shift correction and of the source FWHM)
and to set up the HRI simulator.
To study in detail the off-axis PSF we make use of ray-tracing simulations.  
This allowed us to work with a large number of photons (of the order of millions)
and to get rid of problems related to the aspect solution.
The HRI data concerning the mirror assembly, dimensions, 
incidence angles etc., has been taken from Aschenbach 1988.
Rather than considering details of the mirror surfaces (such as surface roughness, 
correlation length etc.) which mainly affect the PSF width, we imposed the on-axis
PSF (given by David et al. 1998) and extrapolate it to any off-axis angle by means 
of the ray-tracing simulator. This allows us to by-pass problems related to the 
detailed mirror properties.
 
The on-axis PSF can be described by two Gaussians plus an exponential (David 
et al. 1998). Our simulated data allow us to derive a functional form for the PSF
at any off-axis angle. The on-axis description holds for off-axis angles 
$\theta \lesssim 3'$. For larger off-axis angles a better fit is provided by 
Gaussian plus a King profile. The PSF model is therefore given by:
\begin{equation}
\label{eq_psf}
PSF(R)= \left\{ \begin{array}{lr}
A_1\,\exp{(-{{R^2}\over{2\,\sigma_1^2}})}+A_2\,\exp{(-{{R^2}\over{2\,\sigma_2^2}})}+
A_3\,\exp{(-{R\over{\sigma_3}})} & {\rm for \ \theta\lesssim 3'} \\
A_g\,\exp{-({{R^2}\over{2\,\sigma_g^2}})}+K_1\,(1+({R \over {r_{c}}})^2)^{-I_K} &
{\rm for \ \theta\gtrsim 3'}\\ 
\end{array} \right.
\end{equation}
\noindent where the values of the parameters can be found 
in Table \ref{tab_psf}. 

In Fig. \ref{trace} we compare a real image of 
the bright white dwarf HZ 43 taken with the HRI at an off-axis angle of $15'$
with an image simulated with the ray-tracing program. Images are normalised to 
have the same number of counts. We note that the ray-tracing simulation is able to 
reproduce the bright spot as well as the slightly offset extended halo.

\newpage

\newpage

\newpage

\newpage

\newpage

\newpage

\newpage

\newpage

\newpage

\newpage

\begin{figure*}[!htbp]
{\caption{{\bf See the attached tra\_tot.gif file.}
Source detection in the Trapezium field. The four panels show the 
(smoothed) images at rebin 1, 3, 6 and 10, respectively. Circles marks 
X--ray sources. The sizes of the circles is twice the source sigma 
(modeled as a Gaussian).}
\label{tra_tot}}
\end{figure*}

\newpage

\begin{figure*}[!htbp]
\centerline{\psfig{figure=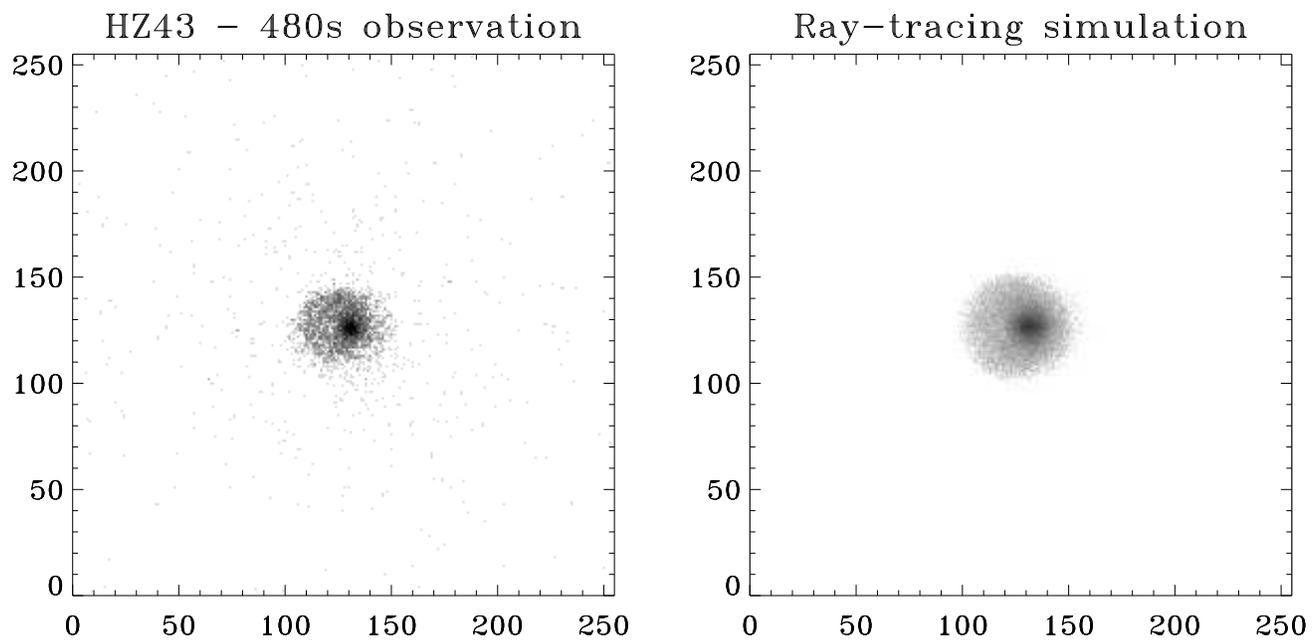,width=18cm}}
{\caption{Comparison between a real $15'$ off-axis HRI image of HZ 43 
(left panel) with a ray-tracing simulation of a source at the same position.}
\label{trace}}
\end{figure*}

\newpage

\begin{table*}[!htbp]
\caption{Correction factor for the nominal count rate.}
\label{cts_off}
\begin{center}
\begin{tabular}{cc}
Off-axis angle ($\theta$)   & Correction ($\alpha$)\\
$\theta\lesssim 2.1'$          & $0.18$ \\
$2.1'\lesssim\theta\lesssim 6.4'$ & $0.14$ \\
$6.4'\lesssim\theta\lesssim 12.8'$& $0.13$ \\
$\theta\gtrsim 12.8'$         & $-0.02+2.04\times10^{-2}\times\theta-7\times10^{-4}
\times\theta^2$ \\
\end{tabular}
\end{center}
\noindent {The nominal count rate $CR$ has to be corrected to account for
the non-Gaussianity of the PSF: the corrected count rate is $CR_{c}=
(1+\alpha)\,CR$. The off-axis angle $\theta$ in the formula for
$\theta\gtrsim 12.8'$ (rebin 10) is in arcmin.}
\end{table*}

\newpage

\begin{table*}[!htbp]
\caption{Comparison between the WDA and other source detection algorithms on 
selected HRI fields.}
\label{fields}
\begin{center}
\renewcommand{\arraystretch}{1.2}
\footnotesize
\begin{tabular}{ccccccccc}
Field name  & Type                & ROR       &Exposure& BMW              & XANADU   & MIDAS \\
            &                     & Number    & (s)    &                  &XIMAGE$^*$& EXSAS \\

\hline
Trapezium  & Star forming region & 200500a00 &\  28089&    222      & 249         &212 \\
           &                     &           &        &(38+82+70+32)&(44+64+73+68)&(34+76+73+29) \\
\hline
47 Tuc     & Globular cluster    & 300059a01 &\  13247&\ 10 	    & \ \ 6       &\ \ 9 \\
	   &                     &           &        &(0+6+2+2)    &(0+3+0+3)    &(0+6+2+1)\\
\hline
M31        & Galaxy              & 150006n00 &\  30790&\   85       & \ 71        &\ 80 \\
           &                     &           &        &(21+40+18+6) &(17+35+12+7) &(21+35+17+7)\\
\hline
NGC 6633   & Open cluster        & 202056a01 &  118806&\ 19         &\ \  5       &\ 17 \\
           &                     &           &        &(1+3+10+5)   &(1+2+1+1)    &(1+6+8+2) \\
\hline
\hline
A 2390     & Cluster of galaxies & 800346n00 &\  27764&\ 15         & \ \ 23       &\ 21 \\
	   &                     &           &        & (3+6+5+1)   & (19+2+2+0)   &(11+5+5+0) \\
\hline
Kepler$^\S$& Supernova Remnant   & 500099n00 &\  36662&\ 51         & 129         &\ 48 \\
\hline
47 Cas$^\S$& Bright X--ray star  & 202057n00 &\  31951&\ \ 1        & \  24       &\ \ 3\\
\end{tabular}
\end{center}
\noindent {$^*$ XIMAGE is not optimised for source detection in ROSAT-HRI 
images so that
a certain number of spurious sources can be found. Moreover, strong sources 
at off-axis angles larger than $\sim 10'$ are often revealed as 
multiple nearby sources.
Source detection has been performed at rebin 1, 3, 6 and 10 
separately, masking the relevant inner region.}

\noindent {$^\S$ Images analysed only at rebin 1. 47 Cas has $\sim 20000$ 
counts (also at 1 spurious source per field, the WDA pick up only one source).}
\end{table*}

\newpage

\begin{table*}[!htbp]
\caption{PSF parameter fits.}
\label{tab_psf}
\begin{center}
\begin{tabular}{cccc}
Parameter & A     & B    & C  \\
$\sigma_1$&\ 1.87 &\,0.04&\,0.20\\ 
$\sigma_2$&\ 3.81 &--0.26 &\,0.30\\
$\sigma_3$&31.11  &\,0.44& 0.00 \\
$A_1\times 10^2$&\ 0.28 &--0.03 & 0.00 \\
$A_2\times 10^2$&\ 7.06 &\,0.79&--0.09\\
$A_3\times 10^4$&\ 2.44 &\,0.23&\,0.02\\
\hline
\hline
$\sigma_g$&\ 2.59&\,0.33&0.04\\
$r_c$     &24.71 &\,6.75&0.35\\
$I_K$     &\ 1.24&\,0.10&0.02\\
\hline
$A_g\times 10^3$&\ 9.28&--2.50 &\,2.34\\
$K_1\times 10^5$&16.63 &\,0.66&--0.05\\
\end{tabular}
\end{center}
\noindent{Angular dependence of the PSF parameters. Parameters have been fit with a
quadratic form $y=A+B\,r+C\,r^2$, with $r$ the off-axis radius in arcmin. 
The normalisations have been chosen such that 
$\int^{\infty}_{0}PSF(r)\,2\,\pi\,r\,dr=1$. Widths are in arcsec.}
\end{table*}

\end{document}